\documentclass{pasa}%

\title[Ultrashort-period eclipsing binaries]{Observations and light curve solutions of ultrashort-period eclipsing binaries}
\author[Kjurkchieva, Dimitrov, Ibryamov and Vasileva]{Diana P. Kjurkchieva$^1$, Dinko P. Dimitrov$^{2,1}$, Sunay I. Ibryamov$^1$ \and Doroteya L. Vasileva$^1$\thanks{based on observations at NAO Rozhen}\\
\affil{$^1$Department of Physics and Astronomy, Shumen University, 9700 Shumen, Bulgaria}%
\affil{$^2$Institute of Astronomy and NAO, Bulgarian Academy of Sciences}}%
\jid{PASA} \doi{10.1017/pas.\the\year.xxx} \jyear{\the\year}

\begin{document}%

\begin{abstract}
Photometric observations in $V$ and $I$ bands and low-dispersion
spectra of ten ultrashort-period binaries (NSVS 2175434, NSVS
2607629, NSVS 5038135, NSVS 8040227, NSVS 9747584, NSVS 4876238,
ASAS 071829-0336.7, SWASP 074658.62+224448.5, NSVS 2729229, NSVS
10632802) are presented. One of them, NSVS 2729229, is newly
discovered target. The results from modeling and analysis of our
observations revealed that: (i) Eight targets have overcontact
configurations with considerable fillout factor (up to 0.5) while
NSVS 4876238 and ASAS 0718-03 have almost contact configurations;
(ii) NSVS 4876238 is rare ultrashort-period binary of detached
type; (iii) all stellar components are late dwarfs; (iv) the
temperature difference of the components of each target does not
exceed 400 K; (v) NSVS 2175434 and SWASP 074658.62+224448.5
exhibit total eclipses and their parameters could be assumed as
well-determined; (v) NSVS 2729229 shows emission in the H$\alpha$
line. Masses, radii and luminosities of the stellar components
were estimated by the empirical relation ''period, orbital axis''
for short- and ultrashort-period binaries. We found linear
relations mass-luminosity and mass-radius for the stellar
components of our targets.
\end{abstract}

\begin{keywords}
(stars:) binaries (including multiple): close; (stars:) binaries: eclipsing;
stars: late-type; stars: fundamental parameters
\end{keywords}

\maketitle%

\section{Introduction}

The investigation of short-period contact binary systems is
important for the modern astrophysics because they are natural
laboratories for study of the late stage of the stellar evolution
connected with the processes of mass and angular momentum loss,
merging or fusion of the stars (Martin et al. 2011). Moreover, the
period-color-luminosity relation makes them useful tracers of
distance on small scales (Rucinski 1994; Klagyivik $\&$ Csizmadia
2004; Eker et al., 2008).

But the statistics of W UMa stars with short and ultrashort
periods ($P \leq$ 0.23 d) is relatively poor (Norton et al. 2011,
Terrell et al. 2012, Lohr et al. 2013) due to a very sharp decline
of the period distribution of binaries with periods below 0.27
days (Drake et al., 2014) as well as faintness of these late
stars.

This paper is continuation of our study of ultrashort-period
binaries (Dimitrov $\&$ Kjurkchieva 2010, 2015, Kjurkchieva et al.
2015, 2016, Kjurkchieva $\&$ Dimitrov 2015). It presents results
of our photometric and low-resolution spectral observations of ten
binaries with orbital periods within 5.1--5.5 hours: NSVS 2175434;
NSVS 2607629; NSVS 5038135; NSVS 8040227; NSVS 9747584; NSVS
4876238; ASAS 071829-0336.7 (further ASAS 0718-03); SWASP
074658.62+224448.5 (further SWASP 0746+22); NSVS 2729229; NSVS
10632802. The targets were selected mainly as a result of
searching for short-period eclipsing binaries (Dimitrov 2009) from
the NSVS database (Wozniak et al. 2004). Additionally, we included
in the sample the known ultrashort-period stars ASAS 0718-03
(Pribulla et al. 2009) and SWASP 0746+22 (Norton et al. 2011) for
follow-up observations.

Table~1 presents the coordinates of our targets and available
information for their light variability.

\begin{table*}
\scriptsize
\begin{center}
\caption[]{Preliminary information about the targets
 \label{t1}}
 \begin{tabular}{cccccccccl}
\hline\hline
Target        & Other name                 & RA          & DEC        & mag       & Ampl     & $P$         & Type & Ref \\
  \hline
NSVS 2175434  &                            & 04 55 22.01 & 64 53 08.6 & 13.65(R1) & 1.02     & 0.220951060 & EW & 1 \\
NSVS 2607629  &                            & 11 42 25.39 & 54 52 48.4 & 12.51(R1) & 0.93     & 0.229370190 & EW & 1, 13 \\
NSVS 5038135  & LINEAR 8209250             & 13 01 11.06 & 42 02 12.7 & 15.34     & 0.35     & 0.2254351   & EW & 3, 4, 12\\
              & 1SWASP J130111.22+420214.0 &             &            &           &          &             &    &  \\
NSVS 8040227  & 1SWASP J173003.21+344509.4 & 17 30 03.06 & 34 45 02.4 & 13.78     & 0.2      & 0.2237144   & EW & 3, 4, 8\\
NSVS 9747584  & 1SWASP J061850.43+220511.9 & 06 18 50.43 & 22 05 11.9 & 13.87(V)  & 0.18     & 0.2143932   & EW & 2, 3, 4 \\

NSVS 4876238  & T-UMa0-03640               & 09 40 57.29 & 51 30 39.0 & 13.76(R)  & 0.242    & 0.22183999  & EC & 11 \\
ASAS 0718-03  & V0989 Mon                  & 07 18 28.67 &-03 36 39.6 & 13.75(V)  & 0.72     & 0.2112594   & EW & 5--9  \\
SWASP 0746+22 & CRTS J074658.6+224448      & 07 46 58.62 & 22 44 48.5 & 14.27     & 0.45     & 0.2208496   & EW & 3, 4, 8  \\
NSVS 2729229  &                            & 13 50 06.20 & 57 26 32.7 & 14.29     &          & 0.22884     & new&  \\
NSVS 10632802 & LINEAR 15208838            & 15 39 51.11 & 10 54 20.4 & 15.36     & 0.6      & 0.2207213   & EW & 3, 12  \\
              & 1SWASP J153951.12+105420.7 &             &            &           &          &             &    &  \\
  \hline
\end{tabular}
\end{center}
Reference: 1 - VSX (Shaw); 2 - Molnar et al. 2013; 3 - Lohr et al.
2013; 4 - Butters et al. 2010; 5 - Kazarovets et al. 2015; 6 -
Pribulla et al. 2009; 7 - Pojmanski 2002; 8 - Norton et al. 2011;
9 - Rucinski 2007; 10 - Drake et al. 2014; 11 - Devor et al. 2008;
12 - Palaversa et al. 2013; 13 - Gurol $\&$ Michel 2017
\end{table*}

\section{Observations}

Our photometric observations of the targets in $V, I$ bands were
carried out in 2010 and 2011 (Table~2) at Rozhen Observatory with
the 60-cm Cassegrain telescope using CCD camera FLI PL09000 (3056
$\times$ 3056 pixels, 12 $\mu$m/pixel) or with 50/70-cm Schmidt
telescope equipped by CCD camera FLI PL16803 (4096 $\times$ 4096
pixels, 9 $\mu$m/pixel). The mean photometric error for each
target and each filter are given in Table 3.

The photometric data were reduced in a standard way with the
software package MaxImDL by dark substraction and flat-field
division. We performed aperture photometry using more than six
comparison stars in the observed field of each target. The
photometric data were phased with the periods from Table 1 and the
corresponding folded curves are presented in Figs. 1--2. The
period of the newly-discovered ultrashort-period binary NSVS
2729229 was determined by the software \emph{PerSea}.

We noted that the amplitudes of half of our light curves differ
from the earlier values (Table 1): those of NSVS 2175434 and NSVS
2607629 are smaller while those of NSVS 8040227 and NSVS 9747584
are bigger (Figs. 1--2). The most amazing case is NSVS 2175434
whose amplitude turned out twice smaller than the earlier value.

\begin{table}
\scriptsize
\begin{center}
\caption[]{Journal of the Rozhen observations \label{t2}}
 \begin{tabular}{cccc}
\hline\hline
Target        & Date        & Exposure [s]& Telescope \\
              &             & $V, I$  &  \\
  \hline
NSVS 2175434  & 2010 Nov 24 & 120, 90  & 60 cm  \\
              & 2010 Nov 27 & 120, 120 & 60 cm  \\
              & 2011 Jan 17 & 120, 120 & 60 cm  \\
              & 2011 Nov 03 & 300      & 2-m \\
           \hline
NSVS 2607629  & 2011 Mar 11 & 30, 60   & 60 cm \\
              & 2011 Mar 12 & 30, 60   & 60 cm \\
              & 2011 Mar 13 & 300      & 2-m \\
 \hline
NSVS 5038135  & 2011 Mar 14 & 90, 90   & Schmidt \\
              & 2011 Apr 11 & 120, 120 & 60 cm  \\
              & 2011 Apr 14 & 120, 120 & 60 cm  \\
              & 2011 Feb 08 & 300      & 2-m  \\
 \hline
NSVS 8040227  & 2011 Aug 06 & 120, 120 & 60 cm  \\
              & 2011 Aug 07 & 120, 120 & 60 cm  \\
              & 2011 Aug 05 & 120, 120 & 60 cm  \\
              & 2011 Mar 13 & 300      & 2-m  \\
 \hline
NSVS 9747584  & 2011 Jan 06 & 120, 120 & 60 cm  \\
              & 2011 Jan 07 & 120, 90  & 60 cm  \\
              & 2011 Mar 11 & 120      & 2-m    \\
        \hline
NSVS 4876238  & 2011 Jan 17 & 120, 120 & 60 cm  \\
              & 2011 Jan 30 & 120, 120 & 60 cm  \\
              & 2011 Mar 11 & 300      & 2-m \\
              \hline
ASAS 0718-03  & 2011 Feb 11 & 120, 120 & Schmidt \\
              & 2011 Mar 11 & 120, 120 & 60 cm   \\
              & 2011 Mar 12 & 120, 120 & 60 cm   \\
   \hline
SWASP 0746+22 & 2011 Apr 11 & 120, 120 & 60 cm  \\
              & 2011 Apr 14 & 120, 120 & 60 cm  \\
              & 2011 Mar 12 & 300      & 2-m \\
  \hline
NSVS 2729229  & 2011 May 06 & 120, 120 &  60 cm  \\
              & 2011 May 07 & 120, 120 &  60 cm  \\
              & 2011 May 21 & 120, 120 &  60 cm  \\
              & 2011 May 22 & 120, 120 &  60 cm  \\
              & 2011 Mar 12 & 300      &  2-m \\
 \hline
NSVS 10632802 & 2011 May 23 & 120, 120 &  60 cm  \\
              & 2011 Mar 13 & 300      &  2-m   \\
\hline
\end{tabular}
\end{center}
\end{table}

\begin{table*}
\scriptsize
\begin{center}
\caption[]{Target spectral type and temperatures $T_m$, and mean
precision $\sigma$ of our VI photometric observations
 \label{t3}}
 \begin{tabular}{ccccccccccc}
\hline\hline
Target          & NSVS     & NSVS   & NSVS    & NSVS    & NSVS    & NSVS    & ASAS    & SWASP   & NSVS    & NSVS \\
                & 2175434  & 2607629& 5038135 & 8040227 & 9747584 & 4876238 & 0718-03 & 0746+22 & 2729229 & 10632802\\
  \hline
Sp type         & K2      & G9     & K4      & K4      & K9      & K9      & K7      & K4      & K9      & K3 \\
$T_{m}$ [K]     & 4900    & 5300   & 4600    & 4600    & 3850    & 3850    & 4000    & 4600    & 3850    & 4750 \\
\hline
$\sigma_V$ [mag]& 0.004   & 0.003  & 0.006   & 0.004   & 0.006   & 0.005   & 0.005   & 0.005   & 0.005   & 0.005 \\
$\sigma_I$ [mag]& 0.003   & 0.002  & 0.004   & 0.003   & 0.003   & 0.003   & 0.004   & 0.003   & 0.003   & 0.005 \\
  \hline
\end{tabular}
\end{center}
\end{table*}

The low-resolution spectral observations were carried out during 5
nights in 2011 (Table 2) by the 2-m RCC telescope of Rozhen
Observatory equipped with focal reducer FoReRo-2 and CCD camera
VersArray (512 $\times$ 512 pixels, 24 $\mu$m/pixel). We used
grism with 300 lines/mm that gives resolution 5.223 A/pixel in the
range 5000-7000 A. The reduction of the spectra was performed with
IRAF package by bias subtraction, flat fielding, cosmic ray and
nebular lines removal and one-dimensional spectrum extraction. The
emission night-sky lines as well as Rb source of emission spectrum
were used for wavelength calibration.

The original photometric data and reduced spectra are
available at $http://astro.shu.bg/obs-data/10stars/$.

The low-resolution spectra (Fig. 3) do not allow radial velocity
measurement and mass ratio determination but are very useful
indicators of temperature and stellar activity. All of the objects
were spectrally classified by the software package \textsc{HAMMER}
of Covey et al. (2007) and the corresponding target temperatures
$T_{m}$ were determined (Table 3). For ASAS 0718-03 we adopted the
photometric temperature from Pribulla et al. (2009) determined by
its 2MASS color index $J-K$.

\begin{figure*}
\begin{center}
\includegraphics[width=14cm,scale=1.00]{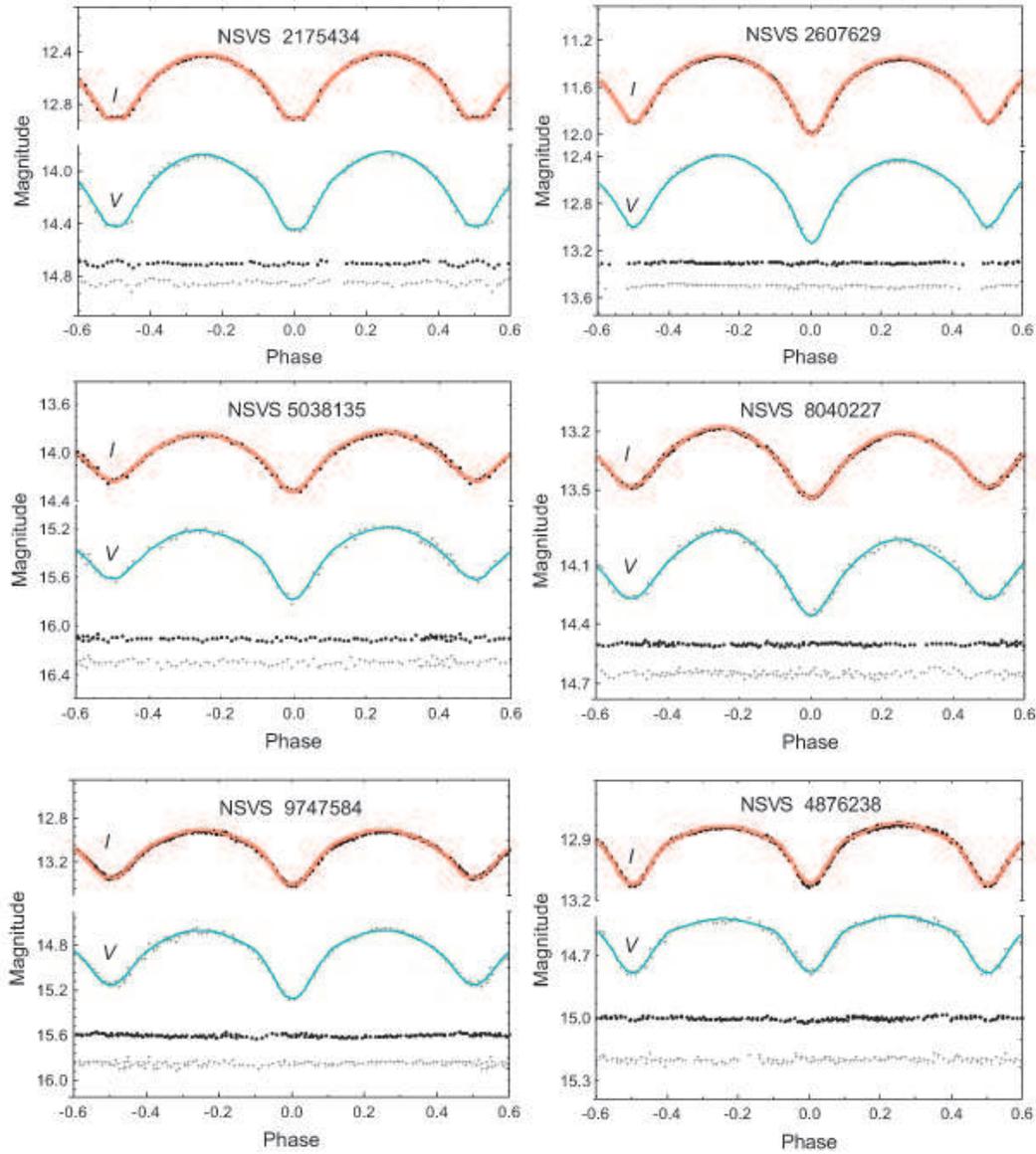}
\caption[]{The folded light curves with their fits and
the corresponding residuals (shifted vertically by different
amount to save space) for the first six targets.} \label{Fig1}
\end{center}
\end{figure*}

\begin{figure*}
\begin{center}
\includegraphics[width=14cm,scale=1.00]{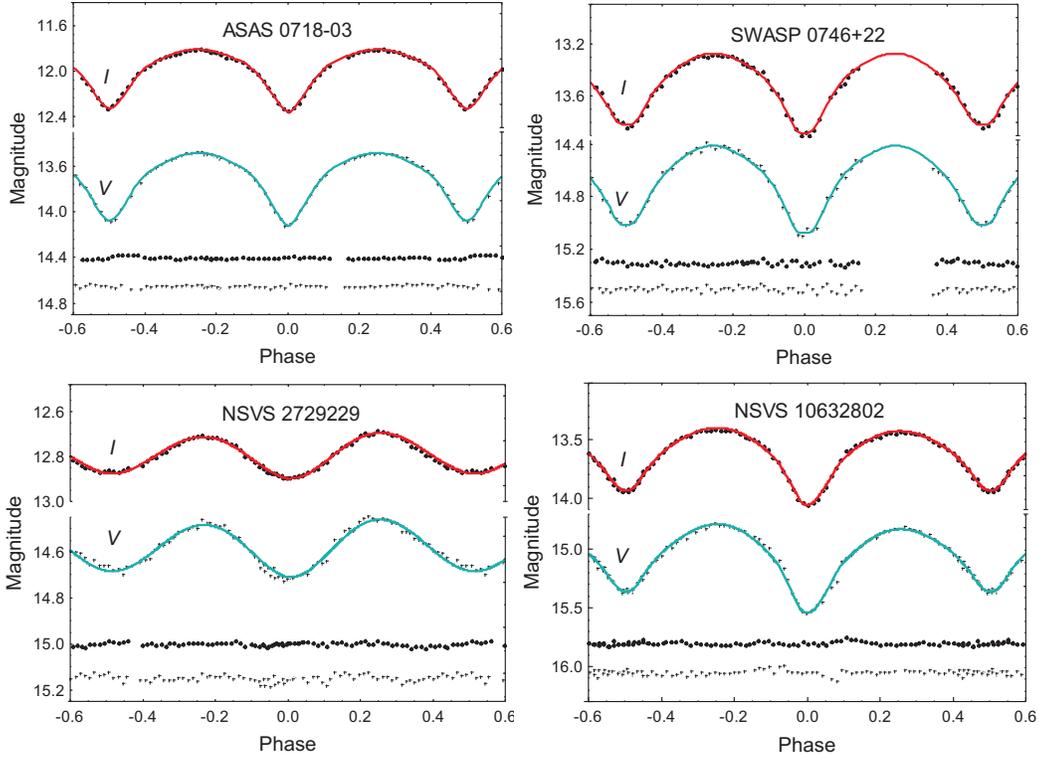}
\caption[]{The same as in Figure 1 for the last four targets.}
\label{Fig12}
\end{center}
\end{figure*}

\begin{figure}
\begin{center}
\includegraphics[width=7.8cm,scale=1.00]{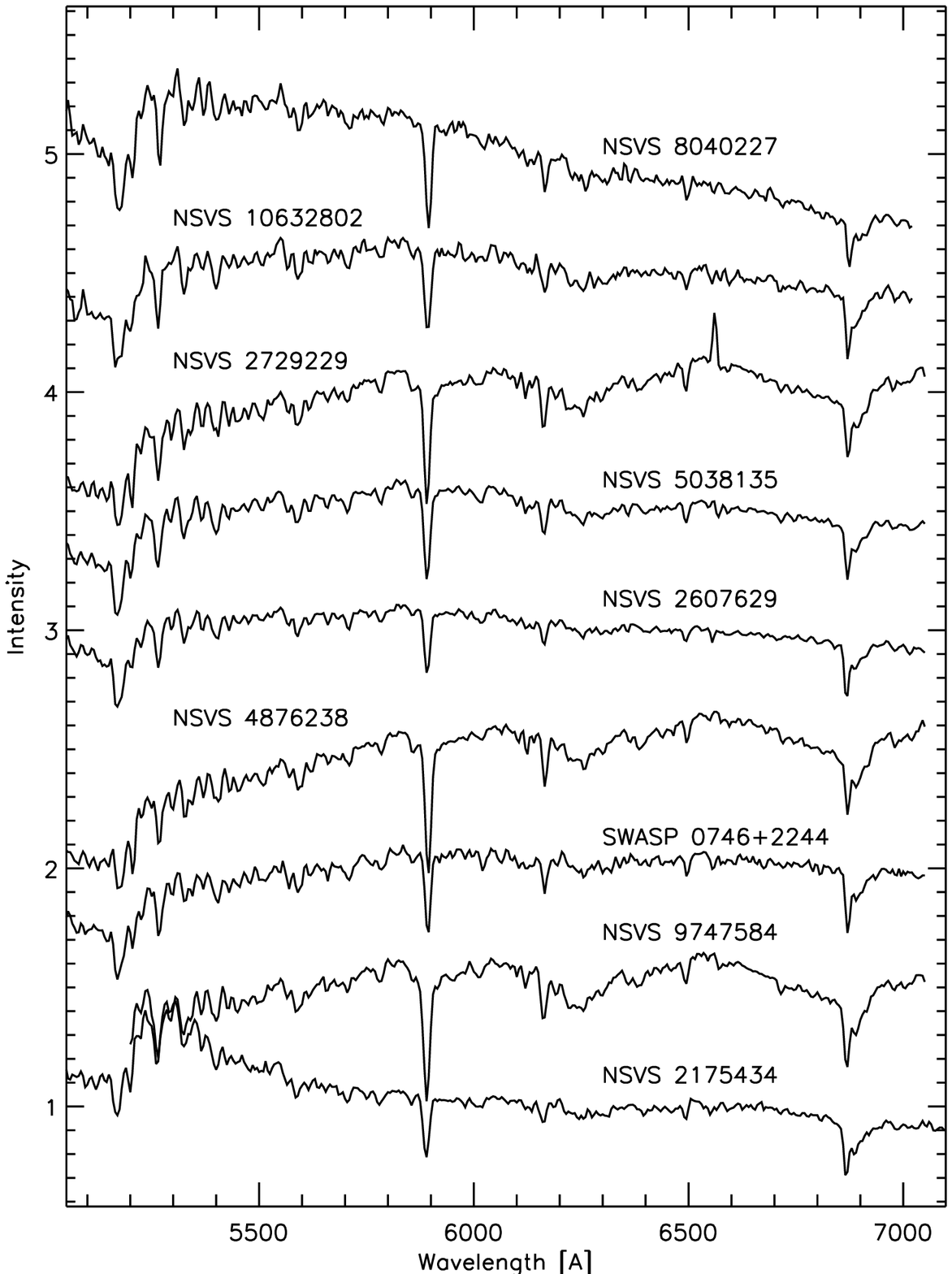}
\caption[]{Low-resolution spectra} \label{Fig2}
\end{center}
\end{figure}

\section{Light curve solutions}

We carried out the modeling of our data by the package
\textsc{PHOEBE} (Prsa $\&$ Zwitter 2005) based on the
Wilson--Devinney code (Wilson $\&$ Devinney 1971).

The procedure of the light curve solutions consists of several
steps. Firstly, we adopted primary temperature $T_{1}$ = $T_{m}$
and assumed that the stellar components are MS stars. Then we
calculated initial (approximate) values of secondary temperature
$T_2$, mass ratio $q$ and ratio of relative stellar radii $k =
r_2/r_1$, based on the empirical relation of MS stars (Ivanov et
al. 2010): $T_2 = T_1 (d_2/d_1)^{1/4}$, $q = (T_2/T_1)^{1.6}$, $k
= q^{0.75}$.

Further we searched for best fit varying: $T_{2}$ and $q$ around
their initial values; orbital inclination $i$ in the range
60-90$^{\circ}$ (appropriate for eclipsing stars); potentials
$\Omega_{1,2}$ in such way that the ratio $r_2/r_1$ to correspond
to its initial value. We adopted coefficients of gravity
brightening 0.32 and reflection effect 0.5 appropriate for late
stars (Table 3). The limb-darkening coefficients were chosen
according to the tables of Van Hamme (1993). In order to reproduce
the light curve distortions we added cool spots on the primary and
varied their parameters (longitude $\lambda$, latitude $\beta$,
angular size $\alpha$ and temperature factor $\kappa$).

After reaching the best solution (corresponding to the minimum of
$\chi^2$) we adjusted the stellar temperatures $T_{1}$ and $T_{2}$
around the value $T_m$ by the formulae (Kjurkchieva $\&$ Vasileva
2015)
\begin{equation}
T_{1}^{f}=T_{\rm {m}} + \frac{c \Delta T}{c+1}
\end{equation}
\begin{equation}
T_{2}^{f}=T_{1}^{f}-\Delta T
\end{equation}
where the quantities $c=L_2/L_1$ (the luminositi ratio of the
stellar components) and $\Delta T=T_{m}-T_{2}$ are determined from
the \textsc{PHOEBE} solution.

\begin{table*}
  \begin{center}
  \caption{Fitted parameters}
  \begin{scriptsize}
  \begin{tabular}{clclllcrc}
    \hline
Star          & $T_0$      & $T_{2}$  & $\Omega_{1}$ & $\Omega_{2}$ & $q$   &  $i$    & $\lambda$ & $\alpha$ \\

 \hline
NSVS 2175434  & 2455579.247328(100) & 4895(19) & 6.32(4)  & 6.32(4)   & 3.01(4)  & 81.9(1) &  90(2)   &  20(1)   \\
NSVS 2607629  & 2455632.779863(50)  & 5090(18) & 2.996(2) & 2.996(2)  & 0.619(3) & 80.2(1) & 270(3)   &  18(1)   \\
NSVS 5038135  & 2455663.518774(200) & 4449(17) & 6.43(2)  & 6.43(2)   & 2.96(1)  & 72.0(1) & 90(2)    &  18(1)   \\
NSVS 8040227  & 2455780.471376(100) & 4356(23) & 2.868(4) & 2.868(4)  & 0.522(3) & 67.2(1) & 270(3)   &  18(1)   \\
NSVS 9747584  & 2455568.511660(100) & 3741(15) & 6.631(8) & 6.631(8)  & 3.091(2) & 77.4(1) &          &          \\

NSVS 4876238  & 2455579.545545(100) & 3816(20) & 2.89(1)  & 2.7(1)    & 0.379(4) & 68.0(1) &  90(3)   &  10(0.5)  \\
ASAS 0718-03  & 2455604.266489(100) & 3942(31) & 3.43(1)  & 3.43(1)   & 0.83(3)  & 76.4(1) &          &           \\
SWASP 0746+22 & 2455663.304142(100) & 4426(32) & 6.09(3)  & 6.09(3)   & 2.84(3)  & 81.7(2) &          &           \\
NSVS 2729229  & 2455688.551888(400) & 3600(20) & 3.32(2)  & 3.32(2)   & 0.81(1)  & 49.6(8) & 110(2)   &  16(1)    \\
NSVS 10632802 & 2455705.435038(200) & 4340(20) & 5.28(4)  & 5.28(4)   & 2.19(1)  & 79.1(1) & 300(3)   &  25(1)    \\
 \hline
  \end{tabular}
  \end{scriptsize}
  \label{table4}
  \end{center}
\end{table*}

\begin{table*}
  \begin{center}
  \caption{Calculated parameters}
  \begin{scriptsize}
  \begin{tabular}{ccccccr}
    \hline
Star          & $T_{1}^{f}$ & $T_{2}^{f}$ & $r_{1}$  & $r_{2}$  & $L_{2}/L_{1}$ & $f$     \\
 \hline
NSVS 2175434  & 4903(20)    & 4898(19)    & 0.321(2) & 0.507(2) & 2.48 & 0.48  \\
NSVS 2607629  & 5390(21)    & 5168(18)    & 0.445(3) & 0.362(3) & 0.55 & 0.28 \\
NSVS 5038135  & 4704(20)    & 4553(17)    & 0.304(1) & 0.491(2) & 2.23 & 0.21  \\
NSVS 8040227  & 4674(26)    & 4430(23)    & 0.449(3) & 0.336(3) & 0.44 & 0.16  \\
NSVS 9747584  & 3947(10)    & 3838(15)    & 0.297(3) & 0.491(3) & 8.13 & 0.17  \\

NSVS 4876238  & 3860(13)    & 3826(12)    & 0.453(1) & 0.264(2) & 0.45 & -0.08  \\
ASAS 0718-03  & 4025(32)    & 3967(31)    & 0.404(2) & 0.371(4) & 0.78 & 0.08  \\
SWASP 0746+22 & 4717(35)    & 4543(32)    & 0.327(1) & 0.505(2) & 2.08 & 0.51  \\
NSVS 2729229  & 3942(27)    & 3692(20)    & 0.421(3) & 0.384(2) & 0.59 & 0.25  \\
NSVS 10632802 & 4986(28)    & 4576(20)    & 0.344(3) & 0.479(3) & 1.36 & 0.39 \\
 \hline
  \end{tabular}
  \end{scriptsize}
  \label{table5}
  \end{center}
\end{table*}

\begin{table*}
  \begin{center}
  \caption{Global parameters}
  \begin{scriptsize}
  \begin{tabular}{ccccccccc}
    \hline
Star          & $a$         & $M_1$       & $M_2$       & $R_1$       & $R_2$     & $L_1$     & $L_2$   & $d$ [pc]  \\
 \hline
NSVS 2175434  &   1.57(5)   &   0.27(3)   &   0.81(7)   &   0.51(2)   &   0.80(4) & 0.131(12) & 0.329(41) & 398(19) \\
NSVS 2607629  &   1.66(5)   &   0.72(6)   &   0.45(4)   &   0.74(3)   &   0.60(4) & 0.412(37) & 0.231(30) & 283(13) \\
NSVS 5038135  &   1.62(5)   &   0.28(3)   &   0.84(8)   &   0.49(2)   &   0.80(4) & 0.106(09) & 0.244(30) & 759(33) \\
NSVS 8040227  &   1.60(5)   &   0.73(7)   &   0.38(4)   &   0.72(3)   &   0.54(3) & 0.222(22) & 0.100(14) & 364(18) \\
NSVS 9747584  &   1.51(5)   &   0.25(2)   &   0.76(8)   &   0.45(2)   &   0.74(4) & 0.044(04) & 0.107(14) & 261(13) \\

NSVS 4876238  &   1.58(5)   &   0.78(8)   &   0.30(3)   &   0.72(2)   &   0.42(2) & 0.102(08) & 0.034(04) & 237(10) \\
ASAS 0718-03  &   1.48(5)   &   0.53(6)   &   0.44(4)   &   0.60(2)   &   0.55(3) & 0.084(09) & 0.067(10) & 249(13) \\
SWASP 0746+22 &   1.57(5)   &   0.28(3)   &   0.79(7)   &   0.52(2)   &   0.80(4) & 0.118(12) & 0.241(33) & 484(24) \\
NSVS 2729229  &   1.65(5)   &   0.64(6)   &   0.52(4)   &   0.70(3)   &   0.64(4) & 0.105(11) & 0.067(09) & 335(17) \\
NSVS 10632802 &   1.57(5)   &   0.34(3)   &   0.74(7)   &   0.54(2)   &   0.75(4) & 0.162(17) & 0.223(29) & 818(43) \\
 \hline
  \end{tabular}
  \end{scriptsize}
  \label{table6}
  \end{center}
\end{table*}

The formal \textsc{PHOEBE} errors of the fitted parameters were
unreasonably small. That is why we estimated the parameter errors
manually by the procedure proposed by Dimitrov et al. (2017).
After reaching the best fit the parameter $b$ was changed around
its value $b^{f}$ (corresponding to the best fit) while the rest
parameters remained fixed until the biggest deviation of the new
synthetic curve from the best-fit synthetic curve became 3$\sigma$
($\sigma$ is the photometric error of the target from Table 3).
Then the difference between the new value of $b$ and $b^{f}$ gives
the precision of this parameter. Figure 4 illustrates our method
for determination of precision of fitted parameters.

\begin{figure}
\begin{center}
\includegraphics[width=8cm,scale=1.00]{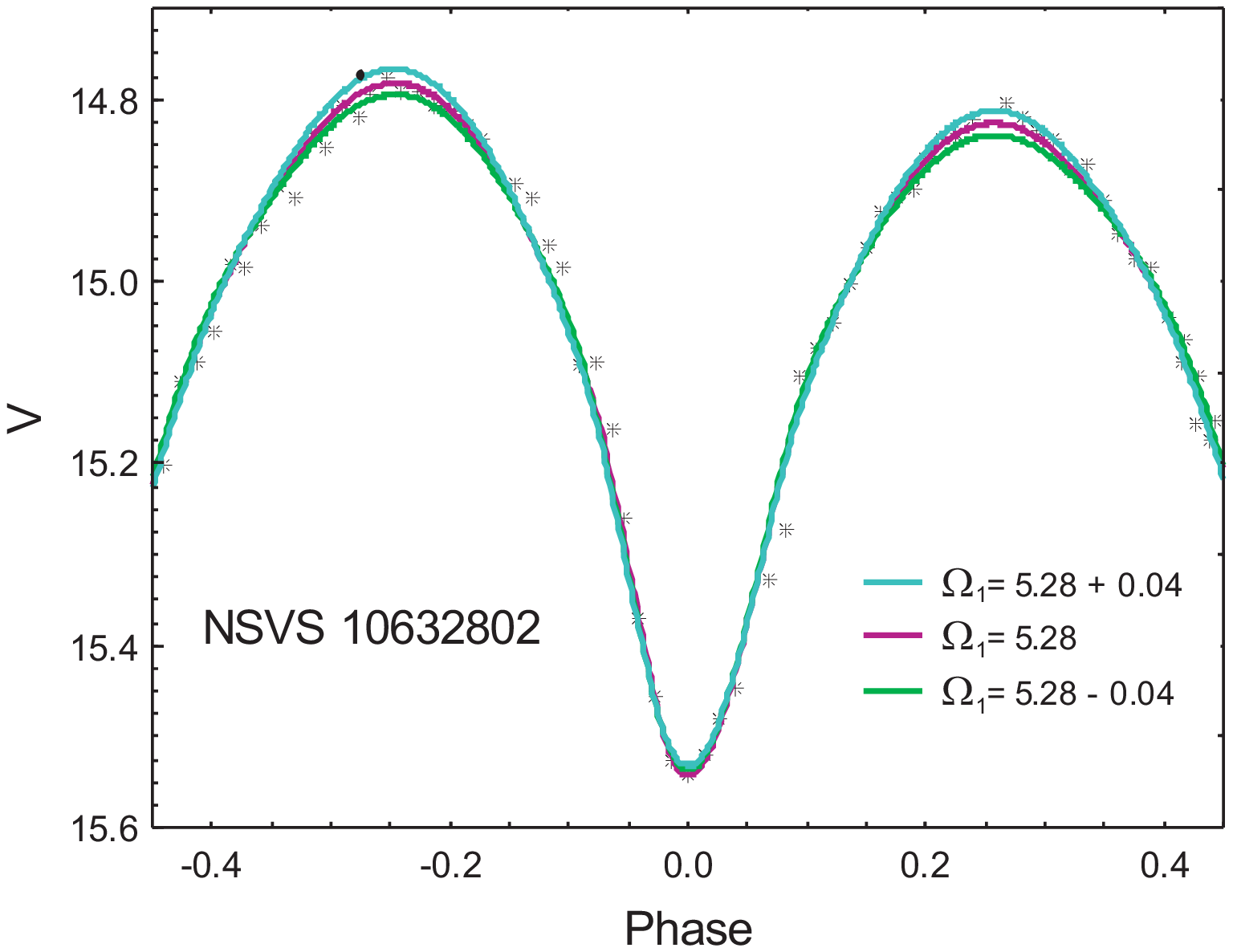}
\caption[]{The three synthetic fits used for determination of the
precision of $\Omega_1$ of NSVS 10632802} \label{Fig5}
\end{center}
\end{figure}

Table~4 contains the final values of the fitted stellar parameters
and their uncertainties: initial epoch $T_0$; secondary
temperature $T_{2}$; potentials $\Omega_{1, 2}$; inclination
\emph{i}; mass ratio \emph{q} and parameters $\lambda$ and
$\alpha$ of equatorial spots with $\kappa =$ 0.9. The synthetic
curves corresponding to the parameters of our light curve
solutions are shown in Figs. 1--2 as continuous lines while Figure
5 exhibits 3D configurations of the targets.

\textsc{PHOEBE} gives a possibility to calculate all values
(polar, point, side, and back) of the relative radius $r_i=R_i/a$
of each component ($R_i$ is linear radius and \emph{a} is orbital
separation). Moreover, \textsc{PHOEBE} yields as output parameters
bolometric magnitudes $M_{bol}^i$ of the two components in
conditional units (when radial velocity data are not available)
but their difference $M_{bol}^2-M_{bol}^1$ determines the true
luminosity ratio $c=L_2/L_1$.

Table 5 exhibits the calculated parameters: stellar temperatures
$T_{1, 2}^f$; relative stellar radii $r_{1, 2}$ (back values);
ratio of stellar luminosities $L_2/L_1$; fillout factor $f$. Using
the empirical relation $P, a$ for short- and ultrashort-period W
UMa type systems (Dimitrov $\&$ Kjurkchieva 2015) we calculated
target orbital axis $a$ (in solar radii) and thus we were able to
estimate masses $M_i$, radii $R_i$ and luminosities $L_i$ of the
stellar components in solar units as well as target distance $d$.
Their values are given in Table 6.

\begin{figure}
\begin{center}
\includegraphics[width=6cm,scale=1.00]{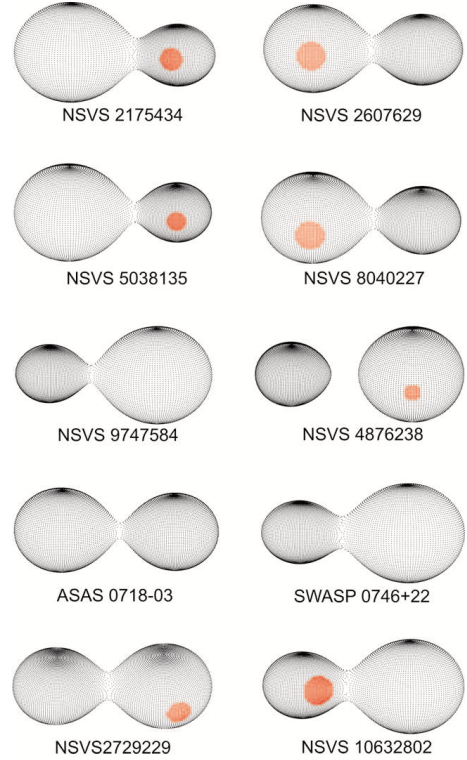}
\caption[]{3D configurations of the targets} \label{Fig4}
\end{center}
\end{figure}

\section{Analysis of the results and discussion}

The main result from the modeling of our photometric and spectral
observations of ten ultrashort-period eclipsing binaries is
determination of: (i) initial epoch $T_0$ of all targets; (ii)
orbital period of NSVS 2729229; (iii) spectral types and
temperatures; (iv) orbital inclination, mass ratio, temperatures
and relative radii of stellar components.

The analysis of the obtained parameter values led to the following
results.

(1) NSVS 2175434 and SWASP 0746+22 exhibit total eclipses and
their photometric mass ratios are reliable (Terrell $\&$
Wilson 2005).

(2) Eight targets (NSVS 2175434, NSVS 2607629, NSVS 5038135, NSVS
8040227, NSVS 9747584, SWASP 0746+22, NSVS 2729229, NSVS 10632802)
have overcontact configurations (Fig. 5, Table 5) with
considerable fillout factor.  NSVS 4876238 and ASAS 0718-03 are
almost contact configuration (with small fillout factor) with
partial eclipses and their photometric mass ratios are worse
determined.

(3) NSVS 4876238 is rare case of detached ultrashort-period
binary, similar to BX Tri (Dimitrov $\&$ Kjurkchieva 2010) and BW3
V38 (Maceroni $\&$ Montalban, 2004).

(4) The temperature differences between the target components are
small (up to 400 K) that is expected for overcontact systems. But
the almost equal component temperatures of the (slightly) detached
binary NSVS 4876238 were not so expected. This result may mean
that NSVS 4876238 is a binary oscillating around a
marginal-contact state predicted by the thermal relaxation theory
(Qian 2002).

(5) All stellar components are K dwarfs excluding the primary of
NSVS 2607629 which is slightly hotter star. The components of NSVS
9747584, NSVS 4876238 and NSVS 2729229 are very late K dwarfs (the
secondary of NSVS 2729229 is rather M dwarf).

(6) Light curve asymmetries of seven targets were reproduced by
cool spots on the side surfaces of their primary components.
Surface inhomogeneities are also appearance of stellar activity of
late stars.

(7) NSVS 2729229 shows emission in H$\alpha$ line (Fig. 3). This
is another appearance (besides the spot) of stellar
activity of this cool star.

(8) We searched for empirical relations of the global stellar
parameters of our ultrashort-period overcontact binaries. It
turned out that there is not well-apparent relation
mass-luminosity (Fig. 5). It may be approximated roughly by the
linear function $L \approx 0.32 M$. In opposite, the relation
mass-radius is quite precise (Fig. 5) and could be described by
the linear function $R = 0.33 + 0.55 M$.

\begin{figure}
\begin{center}
\includegraphics[width=5.8cm,scale=1.00]{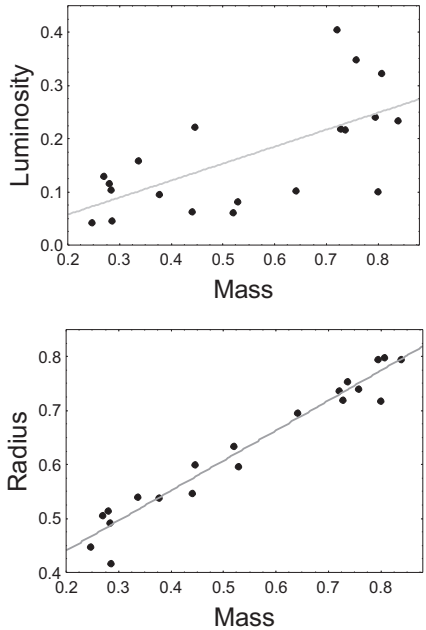}
\caption[]{Relations mass-luminosity and mass-radius for the
target components} \label{Fig5}
\end{center}
\end{figure}

\begin{table*}
\scriptsize
\begin{center}
\caption[]{Two sets of parameters of NSVS 2607629
 \label{t7}}
 \begin{tabular}{ccccc ccccc cccccc}
\hline\hline
Source& $T_1$& $T_2$& $q$  & $i$  & $r_1$& $r_2$& $M_1$& $M_2$& $R_1$& $R_2$& $L_1$& $L_2$& $f$  & $d$ & $\alpha$\\
      & [K]  & [K]  & & [$^{\circ}$]&   &    &[M$\odot$]&[M$\odot$]&[R$\odot$]&[R$\odot$]&[L$\odot$]&[L$\odot$]& & [pc] &[$^{\circ}$]\\
  \hline
G$\&$M& 5420 & 5110 & 1.65 & 77.7 & 0.37 & 0.46 & 0.44 & 0.73 & 0.57 & 0.71 & 0.25 & 0.31 & 0.07 & 144 & 18 \\
our   & 5390 & 5168 & 0.62 & 80.2 & 0.45 & 0.36 & 0.72 & 0.45 & 0.74 & 0.60 & 0.41 & 0.22 & 0.28 & 255 & 59 \\
  \hline
\end{tabular}
\end{center}
\end{table*}

Two of our targets were studied earlier and we present comparison
with our results.

(i) Pribulla et al. (2009) carried out $RI$ photometric
observations of ASAS 0718-03 and obtained estimations of some
parameters: mass ratio $q \sim$ 0.65; orbital inclination $i \sim$
76.8$^{\circ}$; fillout factor $f \sim$ 0. For comparison, by
detailed light curve solution of our $VI$ data we obtained for the
same parameters: $q$ = 0.83; $i$ = 76.4$^{\circ}$; \emph{f} =
0.08. The different values of mass ratio support
insensitivity of the light curve solutions of the
partially-eclipsed binaries to this parameter.

(ii) Based on $BVR_c$ light curves and their modeling Gurol $\&$
Michel (2017) obtained parameters of NSVS 2607629 (first row of
Table 7). The comparison with our values (second row of
Table 7) illustrates the duplicity of the solutions of overcontact
systems: A-subtype configuration corresponding to $q^{(1)} <$ 1
and $r_1^{(1)} > r_2^{(1)}$ and W-subtype configuration
corresponding to $q^{(2)} \approx 1/q^{(1)} >$ 1, $r_1^{(2)}
\approx r_2^{(1)}$ and $r_2^{(2)} \approx r_1^{(1)}$. The
supposition about such W/A ambiguity was made firstly by van Hamme
(1982) and Lapasset $\&$ Claria (1986) who noted that sometimes
both A and W configurations can fit well the photometric
observations and the right choice between the two solutions
requires spectral mass ratio.

Another difference between the two solutions of NSVS 2607629 was
that Gurol $\&$ Michel (2017) used a large hot spot on the
secondary component while we obtained considerably smaller cool
spot on the primary. Although the two solutions reproduce the same
effect (the light at the first quadrature to be smaller than that
at the second quadrature, true for the two runs of observations,
in March 2011 and in March 2016) we assume that cool spots are
physically more reasonable for late stars. This result illustrates
another ambiguity of the light curve solution: hot spot on the
primary has the same effect as appropriate cool spot on the
secondary and viceversa.

\section*{Conclusion}

We presented photometric and low-resolution spectral observations
of ten ultrashort-period eclipsing binaries. The target
temperatures were determined by their spectral type while the
orbital inclination and mass ratio as well as relative radii and
temperatures of stellar components were derived from the light
curve solutions. The stellar masses, radii and luminosities were
estimated by the empirical relation for short- and
ultrashort-period binaries.

This study enriches the poor statistics of ultrashort-period
binaries with known parameters and improved our understanding of
these late stars. Target NSVS 2729229 from our sample of ten
ultrashort-period binaries is newly-discovered member of this
family.

\section*{Acknowledgements}

This study is supported by projects DN 08-20/2016, DN 08-1/2016
and DM 08-02/2016 of National Science Foundation of Bulgarian
Ministry of education and science as well as by projects RD 08-102
and RD 08-80 of Shumen University. The authors are very
grateful to the anonymous referee for the valuable recommendations
and notes.

This research makes use of the SIMBAD database, VizieR service,
and Aladin previewer operated at CDS, Strasbourg, France, and
NASA's Astrophysics Data System Abstract Service.


\end{document}